# INTEGRATING VERTIDROME MANAGEMENT TASKS INTO U-SPACE

Bianca I. Schuchardt (ORCiD), Aditya Devta (ORCiD), Andreas Volkert (ORCiD),
DLR Institute of Flight Guidance, German Aerospace Center, Lilienthalplatz 7, 38108 Braunschweig, Germany

**Abstract**

U-space as defined by the European Commission is a set of new services relying on a high level of digitalization and automation of functions and specific procedures, designed to provide safe, efficient and secure access to airspace for large numbers of unmanned aircraft, operating automatically and even beyond visual line of sight. This kind of concepts of operations (ConOps) of airspace integration for drones and air taxis are also called UTM (Unmanned aircraft system Traffic Management) systems, being U-space the UTM ConOps agreed for Europe. U-space services are under development but commercially not available yet. For demonstration purposes, in the project HorizonUAM, a central U-space cloud service is simulated through a local messaging server using the protocol MQTT (Message Queuing Telemetry Transport). A prototypical vertidrome management tool was created to demonstrate the scheduling and sequencing of air taxi flights. The vertidrome manager is fully integrated within U-space and receives real-time information on flight plans, including requests for start and landing and emergency notifications. Additional information coming from other U-space services (e.g. weather information) can be accessed on request. The integration was demonstrated in a scaled flight test environment with multicopters (<15 kg) representing passenger carrying air taxis.

## NOMENCLATURE

| | |
|---|---|
| AM | Adherence Monitor |
| API | Application Program Interface |
| ATM | Air Traffic Management |
| ConOps | Concept of Operations |
| EMS | Emergency Management Service |
| eVTOL | Electric Vertical Take-off and Landing |
| FATO | Final Approach and Take-off |
| FSA | FATO Safety Area |
| GFMU | Ground Flow Management Unit |
| IoT | Internet of Things |
| JSON | JavaScript Object Notation |
| MQTT | Message Queuing Telemetry Transport |
| QoS | Quality of Service |
| REST | Representational State Transfer |
| RPAS | Remotely Piloted Aircraft Systems |
| TLOF | Touch-down and Lift-off |
| UAM | Urban Air Mobility |
| UAS | Unmanned Aircraft Systems |
| UAV | Unmanned Aerial Vehicle |
| UTM | UAS Traffic Management |
| VATMS | Vertidrome Air Traffic Management System |
| VSO | Vertidrome System Operator |

## 1 INTRODUCTION

"By 2030 drones and their required eco-system will have become an accepted part of the life of EU citizens" as stated by the European Commission in the European Drone Strategy 2.0 [1]. These new airspace users could be smaller drones used for various missions such as for surveillance or cargo delivery. Also air taxis (first piloted, later remotely piloted and finally fully autonomous) could enter the airspace for passenger transport in urban environments. The project HorizonUAM [2] brought together researchers from various disciplines at the German Aerospace Center (DLR) to investigate this Urban Air Mobility (UAM) eco-system holistically with the focus on aerial passenger transport in urban environments. More than 200 cities worldwide have a high potential for the implementation of UAM by the year 2050 according to demand estimations conducted within HorizonUAM [3]. In urban environments, air taxis will most probably not be taking off and landing from conventional runways at existing airports due to capacity limits [4]. It is envisioned that a new type of infrastructure, so called vertidromes or vertiports will be erected [5], [6]. Effective management of these vertidromes is key for catering high-density air taxi traffic.

## 2 U-SPACE FOR AIR TRAFFIC MANAGEMENT

The rapid growth of Unmanned Aircraft Systems (UAS) has introduced new challenges to traditional Air Traffic Management (ATM) systems. To address these challenges, the concept of U-space has emerged, providing a framework for the safe and efficient integration of UAS into airspace systems [7]. While U-space is the European approach, there are other comparable initiatives globally for UAS traffic management (UTM). The FAA UTM ConOps [8] is standing out here. It has many parallels to the U-space approach but still remains rather high-level in technical aspects such as the detailed description of specific UTM services as elaborated in [9].

This section aims to explore the implementation of U-space for air traffic management based on the project CORUS-XUAM (Concept of operations for European UTM systems – Extension for urban air mobility [10]), focusing on its key components, benefits, and potential impact on the aviation industry.

### 2.1 U-space: A Conceptual Framework

U-space refers to a set of technologies, procedures, and regulations that enable safe and efficient operations of drones in low-altitude airspace [11]. It encompasses various aspects of UAS operations, including registration, flight planning, communication, surveillance, and conflict resolution [12]. The concept aims to ensure the integration of UAS into the existing aviation ecosystem, promoting safety, security, and scalability [13]. The key components of U-space are explained in the next section.

### 2.2 Key Components of U-space for Air Traffic Management

The implementation of U-space involves several key components. They are as follows.

#### 2.2.1 UAS Registration and Tracking

UAS operators are required to register their aircraft and obtain a unique identification number [14]. This registration facilitates accountability and traceability, enabling authorities to identify operators and address safety concerns effectively. Moreover, according to EU regulation 2021/664 [14], [15], all aircraft flying in the U-space airspace shall regularly communicate their current position to U-space. As a result, this will form the basis for monitoring, traffic information, tactical conflict prediction and surveillance data exchange [16].

#### 2.2.2 Flight Planning and Management

U-space provides UAS operators with tools for flight planning, including access to airspace information, route optimization, and geofencing capabilities [11]. Through automated systems, operators can submit flight plans and receive real-time information on airspace restrictions and potential conflicts.

#### 2.2.3 Communication and Surveillance

U-space employs a range of communication technologies, including satellite-based systems, to ensure reliable and safe data exchange between UAS, UTM and ground control stations. Surveillance mechanisms, such as ADS-B (Automatic Dependent Surveillance-Broadcast), enable real-time tracking and monitoring of UAS positions, enhancing situational awareness [16].

#### 2.2.4 Conflict Prediction and Resolution

U-space incorporates four methods namely strategic conflict prediction, strategic conflict resolution, tactical conflict prediction and tactical conflict resolution [16] for conflict detection and resolution. Strategic deconfliction is achieved by the flight authorisation service at the time of approving a submitted flight plan which generally happens before the flight [13]. However, tactical deconfliction is performed during the flight. In tactical conflict prediction, alerts are provided to the pilots and the U-space service providers based on current motion and possible intent. On the other hand, tactical conflict resolution utilises sensors and onboard collision-avoidance algorithms to prevent potential conflicts between UAS and other aircraft [11]. These systems support the development of cooperative and non-cooperative collision avoidance capabilities.

### 2.3 Benefits of U-space Implementation

The implementation of U-space offers several benefits to the aviation industry:

- Enhanced Safety and Risk Mitigation

U-space provides a structured framework for UAS operations, ensuring safety through comprehensive registration, flight planning, and surveillance mechanisms [13]. By reducing the risk of mid-air collisions and unauthorized UAS activities, U-space enhances overall airspace safety.

- Increased Efficiency and Scalability

U-space streamlines UAS operations, optimizing flight routes, and minimizing delays. The automation of flight planning and management processes improves operational efficiency, allowing for the scalable integration of a growing number of UAS into airspace [16].

- Facilitation of New Applications and Services

U-space creates opportunities for the development of new UAS applications and services. From drone deliveries to aerial inspections, the implementation of U-space encourages innovation and unlocks the full potential of UAS technology [11].

## 3 VERTIDROME MANAGEMENT

The rapid urbanization and increasing population density in cities around the world have led to ever-growing challenges in transportation infrastructure [17]. Traditional ground-based transportation systems are often plagued by congestion, leading to prolonged commute times, increased carbon emissions, and decreased overall quality of life. To address these issues and embrace the future of transportation, the concept of vertidromes has emerged as a promising solution for UAM. Vertidromes are structures designed to provide spaces for the take-off, landing and maintenance of eVTOL aircraft [18]. As such, they will always provide at least one Touchdown and Lift-off (TLOF) surface, usually a pad, which may additionally be used as a parking space. Figure 1 shows an exemplary layout of a vertidrome and explains its surface features.

The pads are enveloped by the zones for Final Approach and Take-off (FATO) which are extended areas around the pad within which the final approach and initial climb needs to be conducted. To provide an additional layer of safety, these are surrounded by the FATO Safety Area (FSA). TLOF surfaces and parking positions may be connected by taxiways (if the TLOF surface does not also act as a parking space), which has a width that depends on the size of the vehicles expected to operate on the vertiport as well as the type of ground operation (ground taxi vs. hover taxi). As vertidromes emerge as critical infrastructure for UAM, efficient and effective vertidrome management becomes paramount to ensure the safe and seamless operation of Electric Vertical Take-off and Landing (eVTOL) aircraft within urban environments. Vertidrome management involves efficient management of take-off and landing procedures, passenger flow, vehicle charging, and maintenance schedules to minimize waiting times and ensure optimal vehicle utilization [19]. Implementing intelligent algorithms and data-driven analytics for vehicle routing and scheduling can optimize the utilization of vertidrome resources and reduce operational inefficiencies. The information flow for a conceptual Vertidrome Air Traffic Management System (VATMS) is described in the next section.

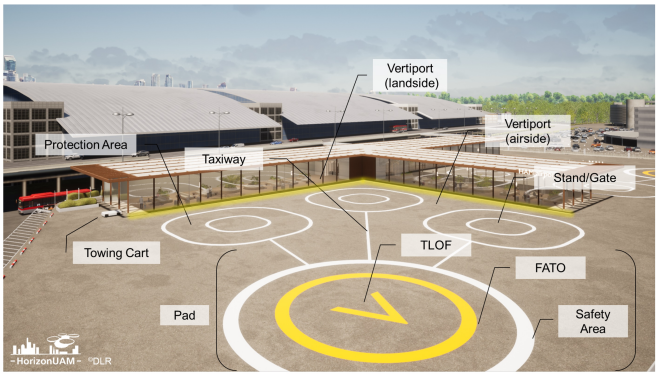

Fig. 1: Concept of a vertidrome [5]

### 3.1 Information Flow

The exchange of data for a conceptual VATMS, developed in the HorizonUAM project, is illustrated in Figure 2.

Entities of the VATMS are represented with boxes, information supplied is contained in hexagons with arrows pointing in the direction of information flow. The input to the system is depicted on the top and left of Figure 2 while the outputs of the system are illustrated on the bottom and right. Additionally, the entities belonging to the area of the vertidrome are represented on the top and bottom of the figure while the U-space cloud services are showed on the left and right.

#### 3.1.1 Inputs to the VATMS

On the input side, the human key component, the Vertidrome System Operator (VSO), who is responsible for overseeing the VATMS, configures operational constraints, such as wind speed limits, as well as (within a prescribed range) thresholds for notifications and alerts of the individual vertiport as to dynamically adjust the operators workload (e.g. suppress notifications of low urgency in emergency situations or during peak hours). The VSO is also authorized to override decisions by the system (e.g. block pads, reclassify a hazard), which also includes decision-making in situations where the automation fails to perform the critical tasks. In Figure 2, the right side of the VSO depicts the avian and drone radar, which provide the location, size and possible trajectories of bird flocks as well as uncooperative vehicles, such as small drones for recreational flying. Those uncooperative airspace users are a risk in the vicinity of airports or vertidromes but also for UAM in general due to the relatively low altitude of operation [20]. Additionally, the Hazard Detection system is illustrated, which relies on a set of sensors and e.g. cameras and image recognition software to supply data, more specifically the location, size and movement, of foreign objects and other (non-environmental) hazards, like ground personnel in FATO vicinity. The left side of the VSO illustrates the local surveillance system, which provides the locations of drones within the vertidrome area of operations in real time. Additionally, a systems monitoring service tracks sensor integrity and infrastructure health, which in this case is serviceability of pads and other equipment directly affecting the capability of vertidrome for accepting arrivals and departures. Information from this service would be shared with and possibly supplemented by ground (maintenance) personnel at the vertidrome itself. The system also receives input from the Ground Flow Management Unit (GFMU), in the form of a pad preference for an arriving or departing vehicle, derived from assigned parking and taxi routes. Arriving and departing aircraft will also transmit their status information to the VATMS and may, in the event of an emergency or loss of connection to the U-space services, establish direct communications with the VATMS and the VSO. Finally, just like at conventional airports, local weather sensors, but also (if necessary) specialized sensors able to detect surrounding urban micro-weather, provide current meteorological data, while forecasts and weather for the greater area are provided by the/a U-space meteorological service. The U-space services communicate with the VATMS through the U-space Cloud Services. The main inputs from the U-space side are provided by the Fleet Manager of the respective vehicles, which are in charge of handling each individual flight. The initial input of the Fleet Manager for a flight is the request for a departure and an arrival slot at the respective vertidromes. During the flight itself (activation of flight plan until end-of-flight reporting) the Fleet Manager provides the flight plan, possible amendments, as well as flight and aircraft status to the vertidrome. The U-space Surveillance Service tracks the flight and supplies information about its location while the Adherence Monitor (AM) verifies the flights adherence to it's flight plan in both time and space. Should the AM detect a discrepancy e.g. a delay, which makes a reserved slot unachievable, it notifies the VATMS and the Fleet Manager. The latter

4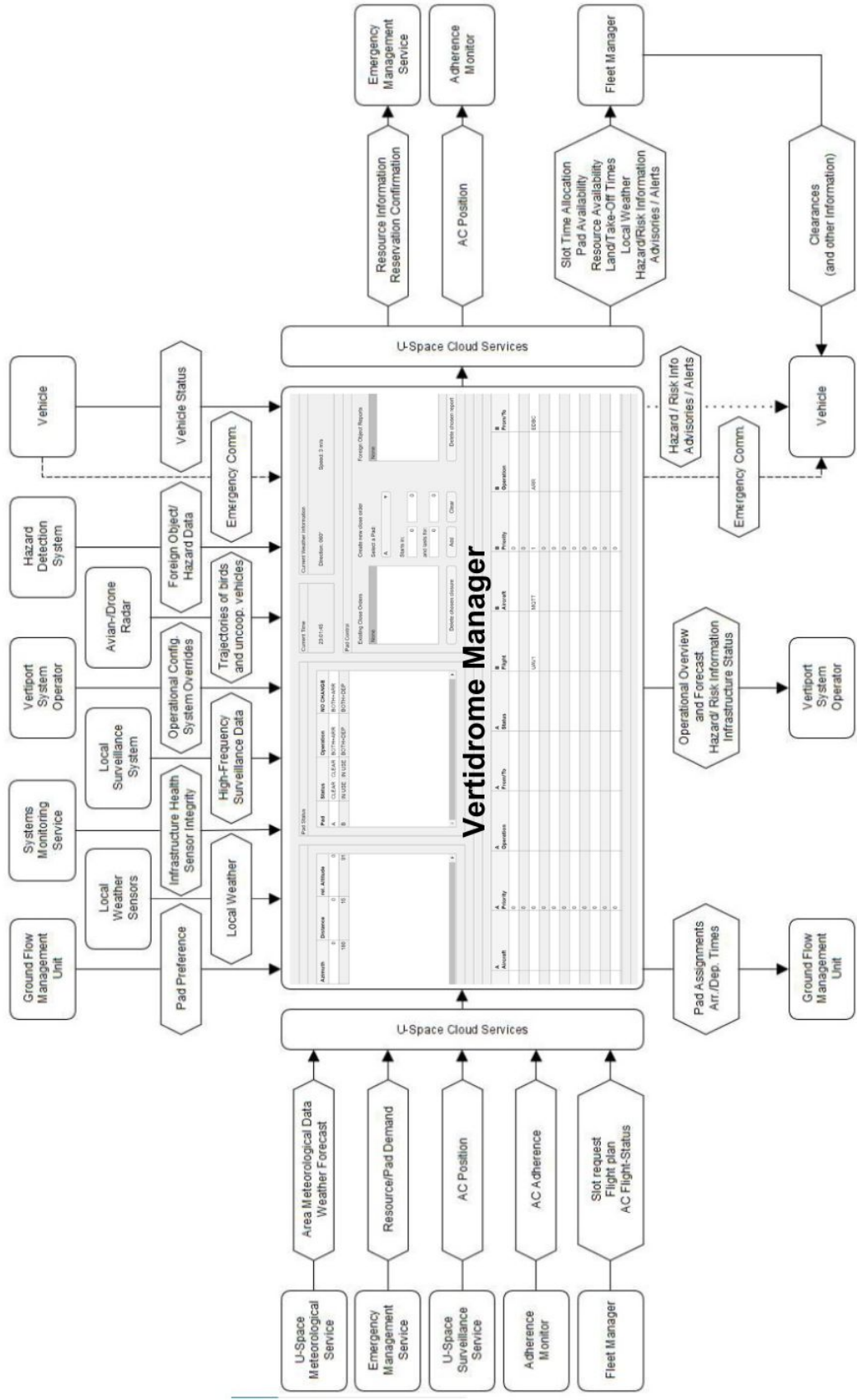

Fig. 2: Information flow of a VATMS

would then have to renegotiate a new slot for the affected vehicle. The remaining input is the U-space Emergency Management Service (EMS). The EMS can, similar to the Fleet Manager, make reservations at a vertiport through the VATMS. However, due to the nature of flights handled by the EMS, these have the highest priority and have to be accommodated by the VATMS, which is why the requests by the EMS are called 'demands' in Figure 2.

### 3.1.2 Outputs of the VATMS

The VATMS generates a wide variety of outputs. The EMS, AM and Fleet Manager receive these outputs through the U-space cloud. The EMS receives confirmation for pad reservations as well as information on available resources/infrastructure that may be utilized to aid the flight in distress. The positions of aircraft are transmitted through to the Adherence Monitor, which will primarily serve the confirmation of punctuality of departing flights and their status regarding entry into their flight corridor. The Fleet Manager receives the confirmed slot times for the planned operation, information on the pad availability and serviceability, as well as information on serviceable resources and infrastructure to assist the aircraft on it's approach or departure. The VATMS also reports observed take off and arrival times, local weather, risks and hazards to the Fleet Manager and submits advisories and alerts about aforementioned observations. Where the environment creates the necessity for better response times, the hazard and risk information, including response advisories and alerts may be directly transmitted to the vehicle. This, however, is supposed to be an optional feed, as indicated by the dotted line. Additionally, the vehicle receives emergency communications (dashed line) in the event the vehicle loses U-space connection on approach or departure. This would include confirmation that a certain pad is reserved for landing, the previously mentioned optional services, as well as establishing a voice communication with the VSO in the case of a human pilot. For clarification, the flow of information and clearances from the Fleet Manager to the vehicle is also illustrated. Under normal conditions, the VATMS does not provide these to the vehicle. After an arrival or departure slot has been confirmed or changed the information is also sent to the Ground Flow Management Unit (GFMU) for planning purposes. Finally, the VSO receives an operational overview and forecast, which includes current and future arrivals and departures, weather information and pad usability. The VSO is also supplied with information on all (potential) risks and hazards and may be prompted to classify a hazard, when the system receives inconclusive or conflicting inputs. The VSO also receives information on sensor integrity and accuracy and overall infrastructure health and availability, which in Figure 2 is summarized as infrastructure health. After explaining about vertidrome management and the information exchange of a conceptual VATMS, the next section addresses the prototypical implementation of such a system for a flight demonstration in a scaled urban scenario in the research project HorizonUAM.

## 4 PROTOTYPICAL IMPLEMENTATION

This section describes the prototypical implementation of the vertidrome manager, the communication protocol and the Fleet Manager.

### 4.1 Vertidrome Manager

Vertidrome Manager is a software prototype in the VATMS ecosystem which is responsible for scheduling, controlling and managing air traffic at a vertidrome. This development is a part of the HorizonUAM project, which researches ways for an efficient, safe and sustainable initial implementation of the U-space SESAR joint undertaking [2]. An overview of the software, which has fully been developed and tested in MATLAB, the data processing and workflows, interface and individual functions, as well as customisation and manual input options are presented in this section. Vertidrome Manager possesses four major subsystems which are the Weather Data Processing Unit, the Adherence Monitoring Unit, the Risk Management Unit and finally the Pad Management and Scheduling Unit. The data flow between these subsystems is shown in Figure 3. The individual subsystems of the software are explained in the subsequent sections.

#### 4.1.1 Weather Data Processing Unit

The Weather Data Processing Unit receives the local weather data from sensors on the vertidrome, as well as the area weather and the forecast for the vertidrome from the Weather Information Service. From this data, the system derives a plan for the pad usability for the Pad Managing and Scheduling Unit, including which pads may be used for arrival, departure or potentially both operations, but also if local weather phenomena require slot times to be extended to allow for safe operations.

#### 4.1.2 Adherence Monitoring Unit

The Adherence Monitoring Unit receives two kinds of input. First, the VSO defines the adherence criteria, which may include tolerances in space and time, as well as define paths that should be followed or areas that shall be avoided for any reason. To be able to ensure that all vehicles comply with these regulations, the system receives high-frequency surveillance data from the appropriate local system. The reason for this data to be from a local source is the requirement for reduced lag and more frequent updates. Within the vertidrome control volume, all aspects of the 4D-trajectory of a vehicle are monitored and in the event of deviations, the Risk Management Unit is supplied with the ID and type of deviation of the respective vehicle. Additionally the VSO is also be alerted about the deviating vehicle to enable quick response and to re-establish compliance/adherence.

#### 4.1.3 Risk Management Unit

The Risk Management Unit is the central system of the VATMS to detect, classify and track any threats to the operation, to alert vehicles of those threats, advise on potential responses and employ mitigation strategies. The Risk Management Unit receives the information about the serviceability of the landing pads, taxiways, landing assistance systems and any infrastructure which may directly or indirectly affect the capability of the vertidrome to safely handle a flight. In case of non-nominal situation, the Risk Management Unit will alert the VSO and notify the U-space stakeholders. Data on foreign objects, personnel within the FATO area or in close proximity and other hazards is also supplied to the unit by the Hazard Detection System. Eventually, the Risk Management Unit processes all the previously mentioned inputs into a set of operational constraints for the Pad Management and Scheduling Unit.

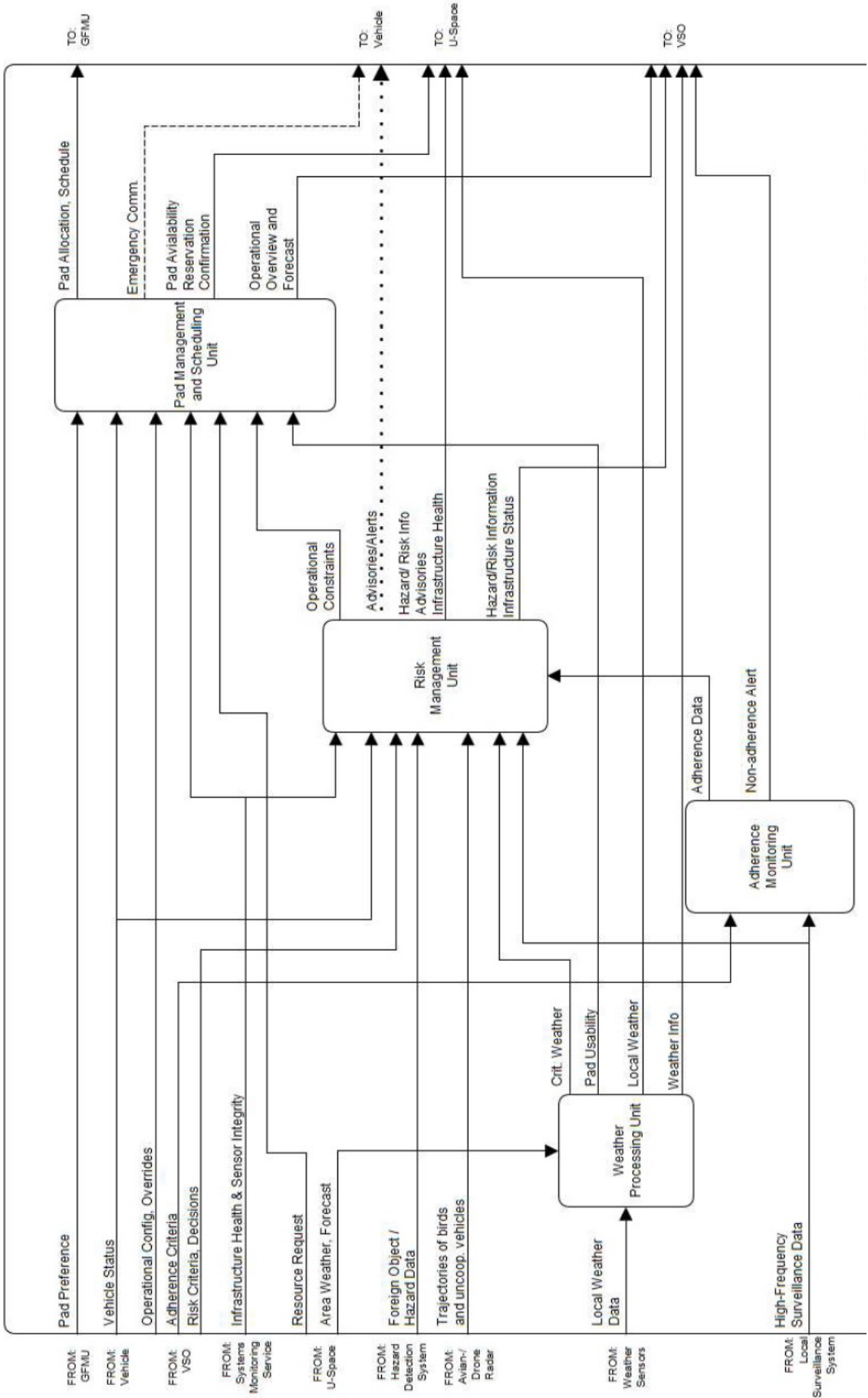

Fig. 3: Internal subsystems of the Vertidrome Manager

### 4.1.4 Pad Management and Scheduling Unit

The Pad Management and Scheduling Unit performs the task of the assignment of FATOs/pads to aircraft requesting operation in and out of the respective vertidrome. Based on the operational constraints imposed by the Risk Management Unit and the Weather Data Processing Unit, the unit confirms a slot for an aircraft requesting landing or departure through a flight plan. All information on planned and in progress flights in and out of the vertidrome, as well as information on the availability and serviceability of pads is compiled into an operational overview and forecast to the VSO. The next section presents the user interface of the Vertidrome Manager.

### 4.1.5 Vertidrome Manager User Interface

The Vertidrome Manager User Interface is illustrated in Figure 4.

In the top left of the figure, a table shows the positions of the aircraft in the pad sectors. The relative heading, distance from the pad and the relative altitude of the incoming or outgoing flight is shown along with its assigned pad. To the right of that is the pad status display which also shows possible changes to the pad usability. Moving to the right, the system clock along with current weather information including wind direction and speed are displayed. Below the weather tab, pad control tab containing displays on pad closures due to foreign objects (right) or orders by the operator (left) as well as a means to create a new close order is shown. The last element on the top is the slot reassignment panel, which allows for flights to be reassigned to empty slots. Finally, on the bottom, an operational forecast is shown, previewing the current scheduled flights for landing or departure from the vertidrome. Additionally, the receival of new flight request from the U-space server is also depicted in Figure 4. As soon as the Vertidrome Manager receives a new flight plan, it notifies the VSO about the flight information as well as about the flight approval or flight rejection through temporary pop-up windows. These pop-up windows also demand an acknowledgement from the VSO in order to make sure the successful exchange of information. After presenting the operator interface of the Vertidrome Manager, the next section explains about the communication protocol of the Vertidrome manager with the U-space server and the Ground Control Station.

## 4.2 Vertidrome Manager Communication Protocol

The Vertidrome Manager uses the MQTT protocol for exchanging data with the U-space server as well as the associated Ground Control Station. MQTT stands for Message Queuing Telemetry Transport. It is a lightweight and widely used messaging protocol designed for efficient communication between devices and applications in the Internet of Things (IoT) and other resource-constrained environments [21]. MQTT was invented by Stanford-Clark and Nipper in the late 1990s [21].

The key features and concepts of MQTT include the following:

- Publish/Subscribe Model: MQTT follows a publish/subscribe messaging pattern, where devices (publishers) send messages to a central broker, and other devices (subscribers) can receive those messages by subscribing to specific topics [22]. Topics act as message channels that organize the communication.

- Lightweight: MQTT is designed to be extremely lightweight, making it suitable for low-bandwidth and high-latency networks [21]. The protocol's minimal overhead reduces data transmission requirements, making it ideal for IoT devices with limited processing power and memory.

- Quality of Service (QoS) Levels: MQTT supports three QoS levels for message delivery. QoS 0 means The message is sent once, and the sender doesn't care if it is received or not. QoS 1 implies the message is guaranteed to be delivered at least once to the receiver with duplicates. QoS 2 means the message is guaranteed to be delivered exactly once to the receiver, without duplicates [23].

- Retained Messages: MQTT allows publishers to set a "retained" flag on messages. When a message is retained, it will be stored on the broker and sent to any new subscribers that join a topic with that retained message [22].

- Last Will and Testament (LWT): Clients can specify a "last will" message that the broker will publish on their behalf if the client unexpectedly disconnects. This feature is useful for notifying others when a device goes offline [23].

- TCP/IP-based: MQTT operates over TCP/IP, but it can also be implemented on top of other transport protocols, such as WebSockets [21].

- Security: MQTT can work with TLS/SSL encryption to ensure secure communication between clients and brokers [21].

The application of MQTT protocol in the case of Ver-tidrome Manager is as follows. Initially, the Vertidrome Manager receives any new flight requests or flight plans from the U-space server over MQTT. According to the calculated results from the Weather Data Processing Unit and the Risk Management Unit, the Vertidrome Manager evaluates the requested pad usability and sends the approval or rejection of the flight plan back to the U-space server over MQTT. The exchange of information over MQTT is also displayed as pop-up windows in the Vertidrome Manager user interface which is illustrated in Figure 4.

Fig. 4: Vertidrome Manager User Interface

Additionally, it can be seen in Figure 4 that as soon as the flight request gets approved, the planned flight is added in the operational forecast tab. The Vertidrome Manager also receives aircraft position coordinates from the U-space server over MQTT. As the aircraft approaches the vertidrome and enters the defined sector, the Vertidrome Manager starts displaying aircraft position information such as distance to the vertidrome, relative altitude and heading to the VSO. The snapshot is depicted in Figure 5.

| Aircraft in Sector | | | | |
|---|---|---|---|---|
| Pad | Flight | Azimuth | Distance | rel. Altitude |
| A | UAV1 | 240 | 15 | 91 |
| B | CLEAR | 0 | 0 | 0 |

Fig. 5: Aircraft position information received by the Vertidrome Manager over MQTT

The Vertidrome Manager also receives information over MQTT about change in pad usability because of detection of personnel or foreign objects on the pad. Reacting to this information, the Vertidrome Manager closes the affected pad for flight operations by generating a close order and displaying the details in the user interface. Figure 6 depicts such a scenario where pad A is closed because of detection of a person.

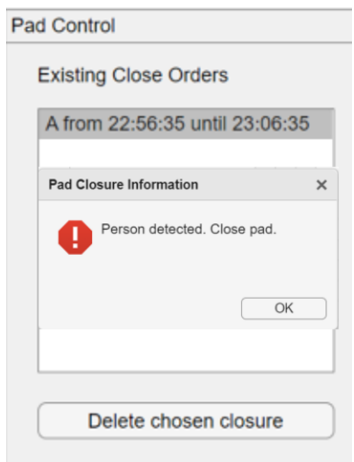

Fig. 6: Pad closure information received over MQTT

After closing the pad, the Vertidrome Manager also forwards this information to the U-space server and the associated Ground Control Station for further steps. As soon as the vertidrome operator clears the pad for operation, this information gets again transmitted over MQTT to the relevant stakeholders.

### 4.3 Fleet Manager

U-Fly is an in-house developed ground control station that has been used for a variety of UAS studies and flight test campaigns such as reported in [24]–[26]. It is used for flight path planning, and execution, as well as operations monitoring. It can be used for the operation of a single or multiple UAS simultaneously. Figure 7 shows the ground operator's view of U-Fly as used for the evaluation of the vertidrome manager. The red areas are no-fly zones, so called geo-fences. The blue circles show the available landing pads. The designated areas for approach or departure are marked as grey arcs. Way points (blue dots) can be adjusted by the UAS operator as needed.

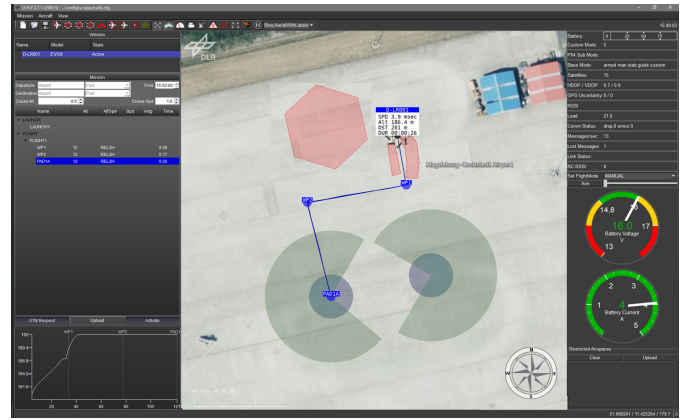

Fig. 7: U-Fly Fleet Manager

Regarding the services as introduced in 2.2 the U-Fly as Fleet Manager assumes the U-space services UAS registration and tracking for displaying the flight path of registered UAS. Flight path planning and management is directly addressed via U-Fly. Communication is established through MQTT as described earlier. The UAS, as introduced in the following section, are currently connected to U-Fly via a 433 MHz telemetry datalink. Strategic conflict prediction and resolution can be performed within U-Fly. For tactical conflict detection and resolution further services onboard the UAS would be required such as described in [27]. This was not part of the vertidrome managager demonstration described in this paper.

## 5 EVALUATION

The vertidrome manager and its integration into a prototypical U-space environment were tested in live demonstrations conducted between May and July 2023 at the National Experimental Test Center for UAS located at Cochstedt, Germany. A model city on a scale of 1:4 was erected from shipping containers [28], see figure 8. Vertidrome landing pads were marked on the ground.

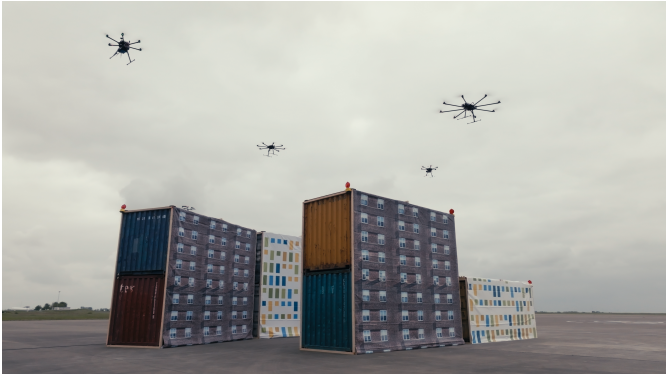

Fig. 8: Urban canyon on a scale of 1:4, erected from shipping containers

### 5.1 Scenario

An airport shuttle use-case was selected for demonstration, similar to a scenario that previously had been evaluated in a virtual reality passenger study [29]. Air taxis are operated between Hamburg 'Airport' and Hamburg city center with a vertidrome named 'Binnenalster'. In the Cochstedt demonstration, an additional landing pad at Hamburg 'Main Station' was included as shown in figure 9. The first demonstration showed the nominal case with one air taxi flying from the North of Hamburg (airport area, behind the model city) above an urban canyon (model city) towards the vertidrome Binnenalster (marked on the ground).

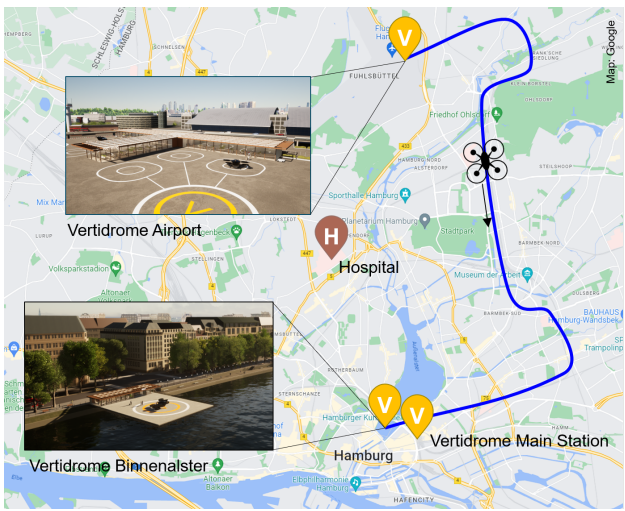

Fig. 9: Urban air mobility scenario within Hamburg

For the nominal procedure the sequence of events was as follows:

1) Fleet manager plans flight path from Airport to Binnenalster within U-Fly
2) Flight plan is sent to U-space for approval
3) Vertidrome manager receives landing request through U-space and accepts it
4) Fleet manager receives approval and initiates take-off at Airport
5) Fleet manager and vertidrome manager can track the air taxi during automated en-route flight
6) Air taxi lands on time at Binnenalster and flight plan is concluded

The second demonstration included a rerouting to the vertidrome Main Station due to a blocked landing pad at Binnenalster. Another UAS detects a passenger on the landing pad Binnenalster, aided by a runtime monitored machine learning algorithm for the detection of persons in image data [30]. For this rerouting scenario the sequence of events is the same as above until step 5 and then continues with:

6) UAS reports detected person on landing pad as emergency through U-space
7) Vertiport manager receives emergency message and closes pad at Binnenalster
8) Fleet manager receives information on pad closure and activates an alternative flight plan to Main Station
9) Vertidrome manager accepts landing request at Main Station
10) Alternative flight plan is approved through U-space and air taxi continues en-route flight
11) Air taxi lands safely at Main Station and flight plan is concluded

Further scaled UAM demonstrations were conducted for the evaluation of drone to drone communication and multisensor navigation. Those results are reported in [31], [32].

### 5.2 Air taxis at scale

For the demonstration within the model city smaller UAS were used to simulate passenger carrying air taxis. An overview of the three drones used for the above scenario shall be provided.

| Figure 10 | EVO X8 heavy |
|---|---|
| manufacturer | Multikopter |
| empty weight | 9.95 kg |
| autopilot | Pixhawk 2.1 orange |
| GPS | Here 2 |
| flight time | > 30 min |
| max wind | 11 m/s |
| max payload | 10 kg |

| Figure 11 | EVO X8 |
|---|---|
| manufacturer | Multikopter |
| empty weight | 7.95 kg |
| autopilot | Pixhawk 2.1 black |
| GPS | Here 2 |
| flight time | > 20 min |
| max wind | 11 m/s |
| max payload | 6.5 kg |

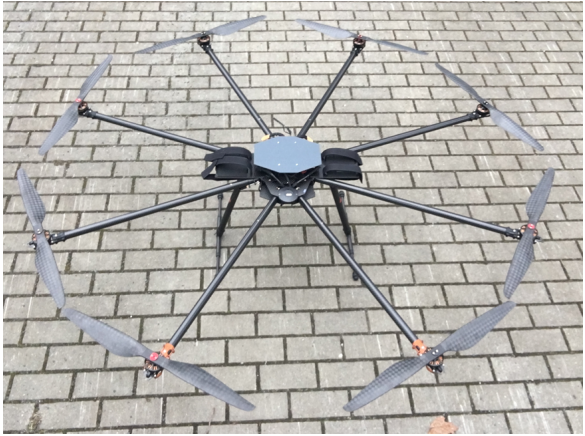

Fig. 10: EVO X8 heavy

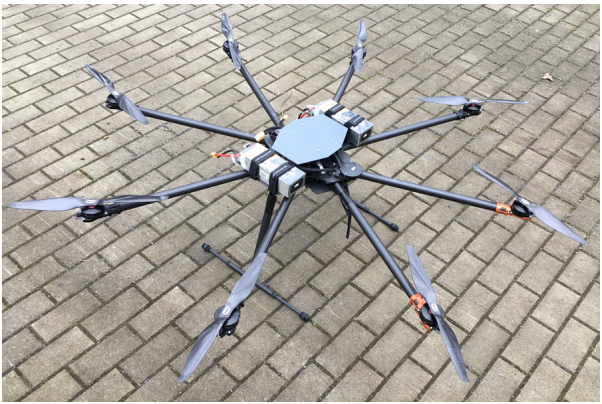

Fig. 11: EVO X8

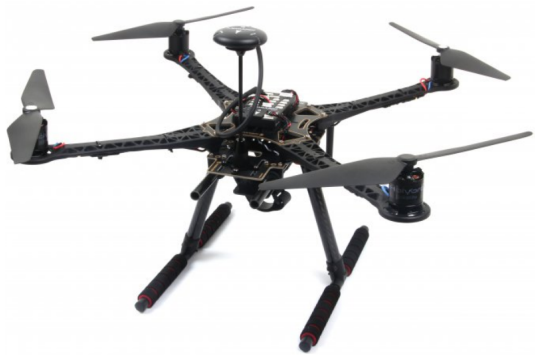

Fig. 12: HolyBro S500 V2

| Figure 12 | HolyBro S500 V2 |
|---|---|
| manufacturer | HolyBro |
| empty weight | 1.3 kg |
| autopilot | Pixhawk 4 |
| GPS | M8N |
| flight time | 15 min |
| max wind | 10 m/s |
| max payload | 1 kg |

For most of the flight trials the EVO X8 heavy [Fig 10] was used, since it fits the scale of the model city of 1:4 the best in order to simulate an air taxi. The EVO X8 [Fig 11] is basically the same drone as the EVO X8 heavy, but with smaller propellers, motors and overall smaller dimensions, thus less empty weight. The HolyBro S500 [Fig 12] was mainly used for preliminary tests and high risk operations. This small drone is easy and cheap to repair in case of an accident, but offers the same autopilot features as the bigger drones.

### 5.3 Results

Figure 13 and 14 show an exemplary flight path from the scaled rerouting scenario, where the incoming air taxi has to divert to an alternative vertidrome. The orange track is the initial flight path originally aiming at vertidrome Binnenalster (rectangular pad marking). The green track shows the then rerouted flight path to Main Station (round pad marking).

The small UAS could successfully be used for scaled demonstrations instead of full-sized air taxis in the trials. In the future the basic assumed U-space services and functionalities are expected to be the same for all UAM vehicles whether passenger carrying or not. The U-Fly interface had previously been used for various experiments and was therefore easily usable also in the air taxi scenarios. The vertidrome manager interface was completely new. It was not optimized for usability yet, but could be used for the intended procedures and commands. Communication via MQTT was successful within the local wireless network. Most difficult to realize in the demonstration was the right timing of events within the short duration of flight. The average nominal scenario had a duration of only 3 minutes with the UAS flying at 2 m/s. Wind was a limiting factor in the tests. During the final runs in July 2023, demonstrations had to be terminated when wind speeds of more than 11 m/s occurred.

## 6 DISCUSSION

The demonstrated vertidrome management tool relies on a human controller to manage incoming requests. Future developments envision a higher degree of automation on the vehicle side but also on the controller side. In future works also the integration at existing airports and the interface with conventional air traffic management will be investigated.

The chosen MQTT protocol proved to be simple to use and effective in mimicking a functional U-space environment. Alternatively, Fas-Millán et al. [33] report the use of a REST-API (Representational State Transfer - Application Program Interface) and JSON (JavaScript Object Notation) format messages in a U-space prototype. Fas-Millán et al. also state that this format allows the easy integration of the different ground control stations and drone platforms, since most programming languages have libraries to manage it; and provides great flexibility to allow changes in the message specifications without necessarily having to impact the existing code. For future ConOps development, an agreed definition of the minimum and optional commands and the information exchanged between the UTM and the UAS would be beneficial [33]. Living labs such as the Air Space Research Area (AREA) U-space [34] currently being implemented in Cochstedt, Germany, will provide a holistic framework for future flight testing within U-space.

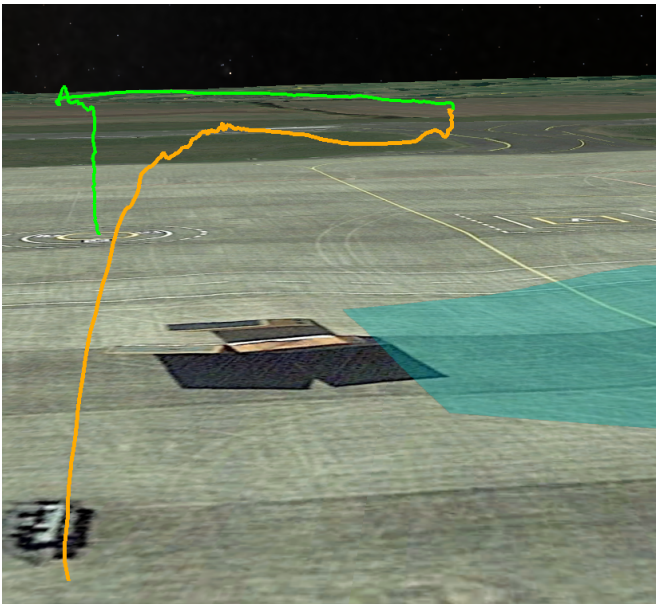

Fig. 13: Flight path of the rerouting scenario: Low view as seen from take-off point

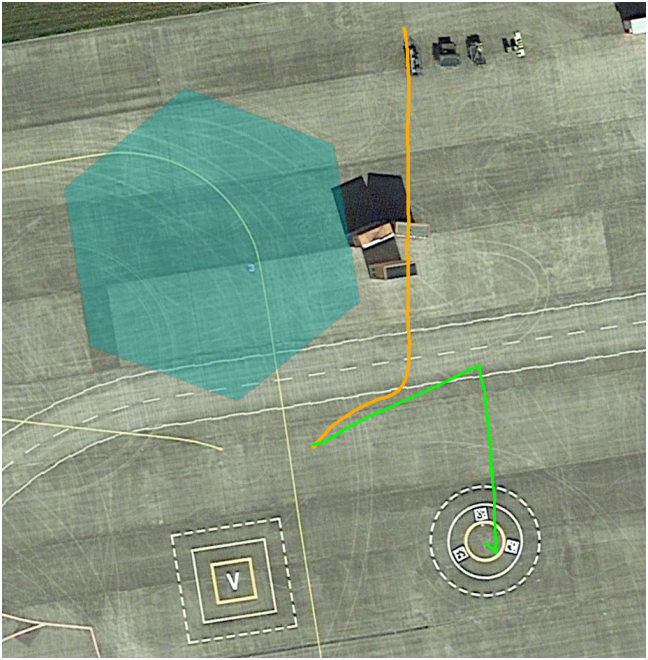

Fig. 14: Flight path of the rerouting scenario: Top view

The vertidrome manager, as conceptualized in this paper, addresses additional U-space services that are not yet described in the CORUS-XUAM ConOps [10]. Required services such as for providing status information on the pad availability or for assistance in emergency procedures will be further investigated in follow-on projects such as EUREKA [35].

## 7 CONCLUSION

This paper reported on the conceptualization, implementation and demonstration of a vertidrome manager. The tool was integrated into a prototypical U-space environment and tested in scaled flight trials with smaller UAS representing air taxis. For this purpose a model city was erected consisting of an urban canyon and several landing pads. The scaled demonstrations proved to be very effective especially as prototypes of full-sized passenger carrying air taxis are still rarely available at the time this paper was written. Specifically, the interface evidenced to be a first good approach to the software that would be required in a first implementation phase in which a person-in-the-loop is required to approve or cancel air taxi operations in terms of functionality and information displayed. Furthermore, the process of the eventualities, even with the requirement of a person involved, was quick enough for the drone in the scaled experiment to be able to maneuver on time an execute the rerouting with no risk.

U-space services will be implemented in the near future in Europe and will be made available for new airspace users such as UAS but also for remotely piloted or autonomous air taxis. Specific U-space services as needed for vertidrome operations were successfully tested in the reported scaled flight demonstrations. In future research, more emphasize should be given on the actual vertidrome layout and the approach procedures. Capacity and efficiency of different designs should be carefully considered in simulation as well as in flight testing.

## ACKNOWLEDGMENT

The authors acknowledge the contribution of Michael Rudolph and Cornelius Lehners in coding the software tools as well as the support of the HorizonUAM flight test team in the validation campaign.

## COMPETING INTERESTS

B.I. Schuchardt is also guest editor for the CEAS Aeronautical Journal for the special issue on the HorizonUAM project but has not been involved in the review of this manuscript. The other co-authors have no competing interests to declare that are relevant to the content of this article.